\documentclass[12pt,preprint]{aastex}

\slugcomment{Submitted to AJ}

\shorttitle{The \textit{BaSTI} database: the AGB extension and Web tools.}
\shortauthors{Cordier et al.}

\begin{document}


\title{A Large Stellar Evolution Database for Population Synthesis Studies. III.
Inclusion of the full Asymptotic Giant Branch phase and Web tools
for stellar population analyses.}


\author{Daniel Cordier}
\affil{Ecole Nationale Superieure de Chimie de Rennes, Campus de Beaulieu, F-35700 Rennes}
\email{daniel.cordier@ensc-rennes.fr}

\author{Adriano Pietrinferni\altaffilmark{1}}
\affil{INAF - Astronomical Observatory of Collurania, via M. Maggini, I-64100 Teramo, Italy}
\email{pietrinferni@oa-teramo.inaf.it}

\author{Santi Cassisi\altaffilmark{2}}
\affil{INAF - Astronomical Observatory of Collurania, via M. Maggini, I-64100 Teramo, Italy}
\email{cassisi@oa-teramo.inaf.it}

\and

\author{Maurizio Salaris\altaffilmark{3}}
\affil{Astrophysics Research Institute, Liverpool John
Moores University, Twelve Quays House, Birkenhead, CH41 1LD, UK}
\email{ms@astro.livjm.ac.uk}


\altaffiltext{1}{Universit\'a di Teramo, viale F. Crucioli I-64100, Teramo, Italy}
\altaffiltext{2}{Instituto de Astrofisica de Canarias, Calle Via Lactea, E38205, La Laguna (Tenerife), Spain}

\altaffiltext{3}{Max Planck Institut f\"ur Astrophysik,
Karl-Schwarzschild-Strasse 1, Garching, D-85741}


\begin{abstract}
Stellar evolution tracks and isochrones are key inputs for a wide range
of astrophysical studies; in particular, they are essential to the 
interpretation of photometric and spectroscopic observations of
resolved and unresolved stellar populations. We have made available to the
astrophysical community a large, homogenous database
of up-to-date stellar tracks and isochrones, and a set of programs useful in
population synthesis studies.  
 In this paper we first summarize the main properties of our stellar model database 
(\textit{BaSTI}) already introduced 
in \citet{pietrin_etal_04} and \citet{pietrin_etal_06}. We then discuss
an important update of the database, i.e., the extension of all stellar models
and isochrones until the end of the thermal pulses along the Asymptotic
Giant Branch. This extension of the
library is particularly relevant for stellar population 
analyses in the near-infrared, or longer wavelengths, where 
the contribution to the integrated photometric properties by cool and
bright Asymptotic Giant Branch stars is significant. A few comparisons
with empirical data are also presentend and briefly discussed.
We then present three \textit{web-tools} that allow an
interactive access to the database, and make possible to compute
user-specified evolutionary tracks, isochrones, stellar luminosity functions,
plus synthetic Color-Magnitude-Diagrams and integrated magnitudes for arbitrary Star 
Formation Histories. All these web tools are available at the \textit{BaSTI} database
official site: \url{http://www.oa-teramo.inaf.it/BASTI}.
\end{abstract}





\section{Introduction} \label{intro}

The interpretation of photometric and spectroscopic observations  
of resolved and unresolved stellar populations is nowadays a fundamental tool to
investigate the formation and evolution of galaxies. Large grids of
stellar evolution models and isochrones are a necessary
ingredient in this kind of analyses, together with
appropriate tools to predict synthetic Color-Magnitude-Diagrams
(CMDs) -- hence star counts along the various observed CMD
branches -- integrated magnitudes and spectra of stellar populations
with an arbitrary Star Formation History (SFH).  
To this purpose \citet{pietrin_etal_04} and \citet{pietrin_etal_06} 
have published a huge and homogeneous database of 
stellar evolution models, that covers the relevant chemical
composition range of stellar populations in galaxies of various
morphological types, and allows one to choose among different treatments
of core convection and stellar mass loss employed in the model
calculations.  
These models are available on the web 
as downloadable files, from the \textit{BaSTI} database -- an acronym for ``\textit{a Bag
of Stellar Tracks and Isochrones}'' -- at
\url{http://www.oa-teramo.inaf.it/BASTI}, together with three
\textit{web-tools} that allow user-friendly manipulations of the
stellar evolution library. Thanks to these tools one can compute
user-specified isochrones, interpolate among evolutionary
tracks to obtain the evolution for a mass value not contained in the grid, 
determine the differential and cumulative luminosity functions from a set of
isochrones and, finally, compute synthetic stellar
populations with an arbitrary SFH and determine their integrated
magnitudes and colors.

This paper describes comprehensively the \textit{BaSTI} database, and in
particular how to run the various \textit{web-tools}, the necessary
inputs and their outputs. Next section describes the stellar
evolution library, with particular emphasis on the most recent
extensions not included in \citet{pietrin_etal_04} and
\citet{pietrin_etal_06}; the following three sections describe 
the individual \textit{web-tools} that allow compute user specified isochrones
and models, luminosity functions and synthetic CMDs.
Final conclusions and an appendix with some more technical details
about the implementation of the \textit{web-tools} close the paper.

%---------------------------------------------------------------------------
\section{Stellar evolution models and isochrones} \label{models}

We have computed stellar evolution models for 11 different metallicities, namely 
$Z=0.0001$, 0.0003, 0.0006, 0.001, 0.002, 0.004, 0.008, 0.01, 0.0198,  
0.03 and 0.04, assuming two different heavy element distributions: 
the scaled-solar one by \citet{Grevesse93}, and the $\alpha$-enhanced
one with $<[\alpha/Fe]>=0.40$ by \citet{salaris_weiss_98}. As for the
initial He-mass fractions, 
we employed a value $Y=0.245$ for the cosmological He (see, e.g.,
\cite{cassisi_etal_03}). To reproduce the solar initial He-abundance obtained
from the calibration of the solar model, we assume an  
Helium enrichment law equal to $\Delta{Y}/\Delta{Z}\sim1.4$. 
For each initial chemical composition we have computed stellar
evolutionary tracks with mass in the  
range $0.5\le{M/M_\odot}\le10$ and a fine mass spacing. 
Mass loss from the stellar surface
is accounted for using the \citet{reimers_75} law and two values of the
free parameter $\eta$, namely $\eta$=0.2 and 0.4. For stars that
develop convective cores during the central H-burning phase we have computed
models with and without overshoot from the Schwarzschild boundary
of the central convective regions.

All evolutions, with the exception of the least massive stars whose central H-burning timescale 
is longer than the Hubble time, have been computed from the Pre-Main Sequence phase until to  
the C-ignition, or until the first thermal pulse along the Asymptotic
Giant Branch (AGB), depending on the initial stellar mass. Full details of these calculations can be 
found in \citet{pietrin_etal_04} and
\citet{pietrin_etal_06}. Very recently we have extended the evolution of
low- and intermediate-mass objects through the entire Thermal Pulse (TP)
phase, until the end of the AGB stage. To this purpose we have used
the synthetic AGB technique (\citet{iben_truran_78}). Our
simplified treatment of the TP-AGB phase reproduces satisfactorily
several integrated  properties (see below) of stellar populations in the near-IR 
bands, which are greatly affected by the presence of TP stars. Given
that the purpose of this library of stellar models is to provide a
reliable tool to be employed in stellar population synthesis studies,
we feel confident that the following simplified treatment of the AGB evolution
is adequate for our purposes.
For each stellar model of a given initial chemical composition and
mass, we started the synthetic AGB evolution at the beginning of the 
TP phase, where the stellar evolution calculations were terminated. 
The first model of the synthetic TP-AGB evolution is
characterized by the total mass $M$, Carbon-Oxygen core mass
$M_{CO}$, 
luminosity $L$, effective temperature $T_{eff}$ and surface chemical
composition ($X_f,Y_f,Z_f$) of the last fully evolutionary model.

The TP-AGB phase is then followed by increasing (after a given
timestep d$t$) $M_{CO}$ and $L$ according to Eqs.~5-7 in 
\citet{wagenhuber_groenewegen_98} which contain a term
mimicking the effect of the Hot Bottom Burning, when appropriate. The
hydrogen mass fraction in the envelope (an input of Eq.~6 of \cite{wagenhuber_groenewegen_98}) is
approximated as $1-(Y_f+0.01)-Z_f$ all along the TP evolution. During
a timestep the mass of the envelope ($M-M_{CO}$) is reduced not only
by the growth of $M_{CO}$, but also by mass loss
processes, accounted for using the mass loss formulae by \citet{vassiliadis_wood_93}. 
For any given value of $M$ and $M_{CO}$, the
effective temperatures are computed using the relationships 
in \citet{wagenhuber_96}, that are plotted in Fig.~8 of 
\citet{wagenhuber_groenewegen_98}. To
ensure continuity, the zero points of the equations describing the
evolution of $L$, $M_{CO}$  and $T_{eff}$ have been adjusted to reproduce the
corresponding values of the last fully evolutionary model, at the
beginning of the TP phase.
The synthetic evolution is stopped when the models have 
started to evolve off the AGB, at constant luminosity, towards their White Dwarf
cooling sequence.
The full models have been used to compute isochrones for ages ranging
between 30~Myr and 19~Gyr. We provide separately models and 
isochrones with and without the inclusion of the TP-AGB phase, 
in case users wish to use their own description of the
TP-AGB evolution, either employing fully evolutionary calculations, or
synthetic AGB models with different choices about, i.e., mass loss
and/or core mass-luminosity relationship. To this purpose we provide
-- as mentioned below -- tables with He and CO core mass 
as well as the luminosity and effective
temperature at the beginning of the TP-AGB phase.

The reader is referred to
\citet{pietrin_etal_04} for a detailed discussion about the
normalization of the individual evolutionary tracks (the concept of
key-point and normalized points between two successive key-points) 
and isochrone computation. 
Here we simply mention two changes with respect to
the discussion in \citet{pietrin_etal_04}. The first change is the
inclusion of two additional key-points along the Red Giant Branch (RGB) one to mark the
maximum luminosity during the RGB bump, and another one to mark the
minimum RGB bump luminosity (key points 6 and 7). A total of 370
normalized points is distributed between key point 6 and its
predecessor, 30 points are distributed between key-points 6 and 7, and
400 points are distributed between key-point 7 and the following one, 
the tip of the RGB. Due to the TP-AGB extension, we have added an
additional key-point (number 17) after the beginning of the TP phase,
to mark the termination of the AGB phase, when the model evolution
turns to the blue. A total of 250 normalized
points cover the TP-AGB phase, between the key-points 16 and 17.
The resulting isochrones have a total of 2250 points.

The main characteristics of the stellar model database are summarized in
Tables~1 and 2. Table~1 lists the grid of initial chemical
compositions; Table~2 summarizes the available sets of models and isochrones.
For a given heavy elements mixture (either scaled-solar or
$\alpha$-enhanced) we have computed models for 11 $Z$ values 
(Table~1) using both $\eta$=0.2 and
$\eta$=0.4 for the evolution until the TP-AGB phase. For each $\eta$
we have computed models with and without overshoot from the central
convective core (in the appropriate mass range). The extension of the
overshooting region $\lambda_{OV}$ is 0.2$H_p$ ($H_p$ is the local
pressure scale height at the Schwarzschild convective boundary)
decreasing to zero when the convective cores vanish, as described in 
\citet{pietrin_etal_04}. For each choice of the metal mixture, $Z$,
$\eta$ and convective core extension we specify the number of tracks
available, their mass range, the number of isochrones available and
their age range. Notice that for masses above $\sim2.4M_{\odot}$ the
evolution with $\eta$=0.4 is identical to the case of $\eta$=0.2
(because of the negligible amount of mass lost during their evolution, 
according to the Reimers' law).

Broadband magnitudes and colors of the stellar evolution tracks and
isochrones are predicted using color-$T_{eff}$ transformations and bolometric corrections 
based on an updated set of model atmospheres described in 
\citet{pietrin_etal_04} and \citet{cassisi_etal_04}. The evolutionary results are available in the
photometric filters listed in Table~2. Additional filters will be
added with time. 
It is worth to point out that, for the first time, we have 
homogenoeus transformations for both scaled-solar and $\alpha-$enhanced
mixture even for super-solar metallicities. At present, the 
transformations for the ACS HST filters are available only for a
scaled-solar chemical composition. 

For the TP-AGB section of the models and isochrones 
broadband colors and magnitudes have
been computed by supplementing the transformations used up to the
beginning of the TP phase with the \citet{westera_etal_02} ones for RGB 
and AGB stars with $T_{eff} <$
3750~K. To ensure continuity, the bolometric corrections and colors of
\citet{westera_etal_02} have been shifted to match the other sets of
transformations when $T_{eff}$=3750~K.
Along the TP-AGB phase, when the $(J-K)$ colors reach $(J-K)$=1.2~mag, we
used the $(J-K)-T_{eff}$ and $(H-K)-T_{eff}$ relationships by 
\citet{bergeat_etal_01} that are appropriate in the carbon star
regime. Models and isochrones extended along the TP-AGB phase are at the
moment transformed only to the $UBVRIJHKL$ system.

All models described above have been
computed in the assumption that a star of a given mass and initial
chemical composition loses during its evolution a fixed amount of
mass, determined by the adopted mass loss law. An isochrone for an old 
population will therefore display on the CMD a clumpy Horizontal
Branch (HB) phase, 
made of stars all essentially with the same total mass, because of
both the constant mass of their progenitor, and the constant value of the mass
lost along the previous RGB phase.   
If one wants to simulate the extended HBs observed in Galactic
globular clusters, a spread in the amount of mass lost by the RGB
progenitor has to be accounted for. To this purpose, for each chemical
composition we have also computed additional large sets of core He-burning
models, with He core mass and envelope chemical profile fixed by a 
RGB progenitor having an age of $\sim13$~Gyr at the RGB tip,
and a range of values of the total stellar mass, to simulate the
effect of a spread in the mass lost along the RGB.
These Horizontal Branch (HB) models ($\sim30$ for each chemical
composition) constitute a valuable tool to perform synthetic HB
modeling, and to investigate pulsational and evolutionary properties
of different kinds of variable stars. 
 
Users can directly download from the \textit{BaSTI} database
evolutionary tracks and HB models as single files, or as tar gzipped 
archive files, selecting among the
following options (11 initial chemical compositions for each choice): 
scaled-solar, $\alpha$-enhanced, canonical (no-overshoot),
non-canonical (overshoot included), $\eta$=0.2, $\eta$=0.4, TP-AGB
excluded, TP-AGB included. We have included in the web page appropriate readme
files and provide also tables (both in ascii and html format)
summarizing the main properties of the theoretical models (not including the
TP-AGB phase) and providing also the CMD location of  
the Zero Age HB (ZAHB) and the central He exhaustion stages. These two
latter sets of tables can be found at the pages containing the
$\eta$=0.4 models. 
A large number of precomputed isochrones for each of the individual options described
before -- spanning the full age range allowed by the computed models
-- can be downloaded as tar gzipped archive files.

The models and isochrones contained in \textit{BaSTI} have been
extensively tested in \citet{pietrin_etal_04} and
\citet{pietrin_etal_06}, and already employed in a number of
investigations (e.g., \cite{gallart_etal_05}). 
We conclude this section showing a few tests for the
TP-AGB extension we introduced in this paper. We have first tested our
isochrones against near-IR Surface-Brightness-Fluctuation (SBF)
magnitudes, that are very sensitive to the brightness and evolutionary
timescales of AGB stars (see, e.g., \cite{liu_etal_00},
\cite{cantiello_etal_03}, \cite{mouhcine_etal_05}, \cite{raimondo_etal_05}). 

We considered the $J$ and $K$-band SBF
magnitudes determined by \citet{gonzalez_etal_04} for a sample of Magellanic
Clouds' superclusters. A supercluster is obtained by coadding
individual clusters belonging to the same SWB class (see
\cite{searle_etal_80} for the definition of the SWB classes, that are
essentially age ranges). \citet{gonzalez_etal_04} provide also the
mean metallicity and age of the objects grouped in each
supercluster. The original SBF magnitudes were on the 2MASS system
and have been transformed to the Johnson system following 
\citet{bessell_brett_88} and \citet{carpenter_01}.

\begin{figure*}
\epsscale{1}
\plotone{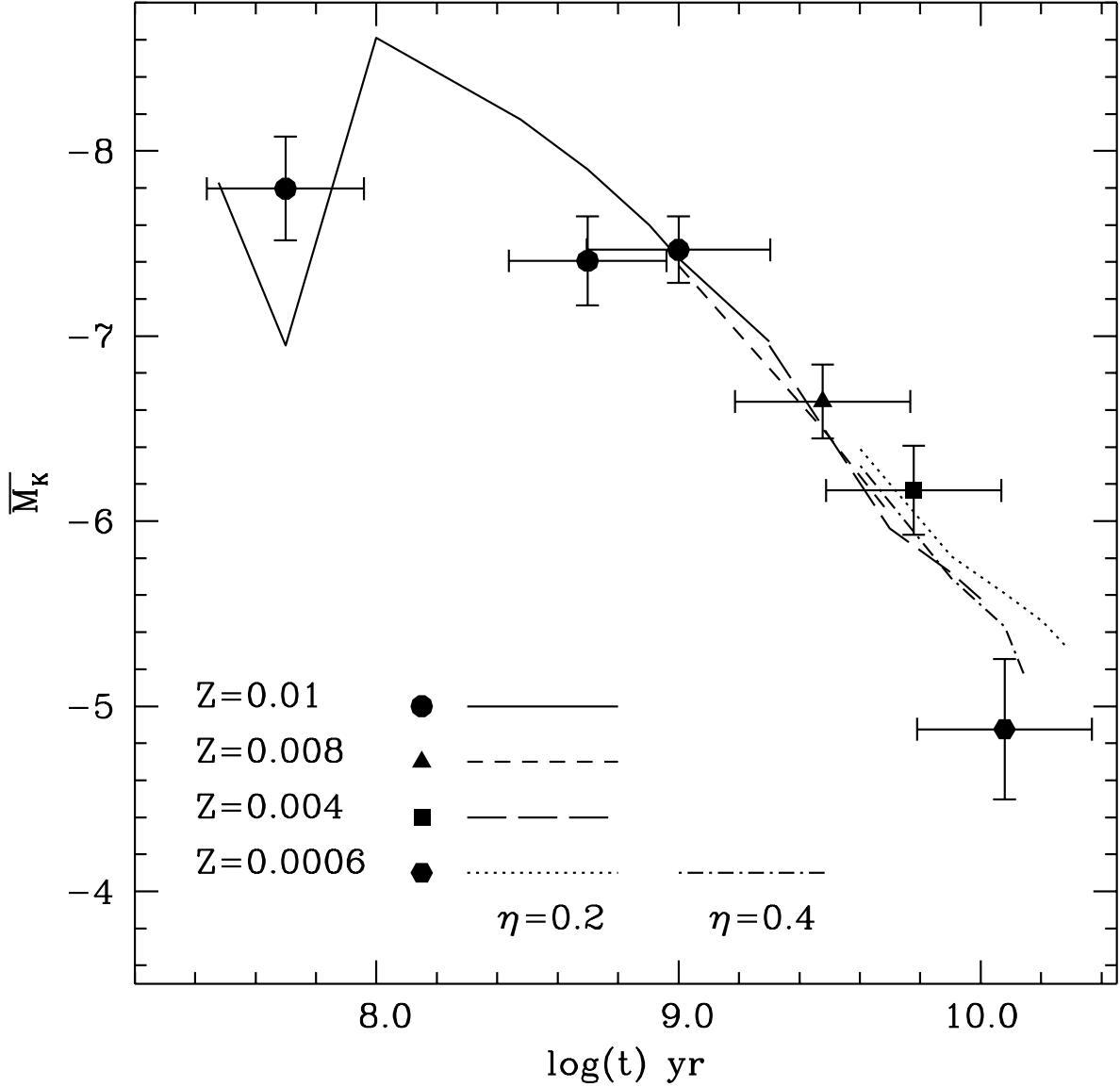}
\caption{Run of the $K$-band SBF magnitudes of Magellanic Cloud
superclusters as a function of their age.
Theoretical SBF magnitudes for the appropriate 
metallicity and age range (denoted with different line styles) 
of the individual superclusters
(superclusters of different metallicities are denoted with different
symbols) are also displayed. 
These theoretical values are obtained from the scaled-solar isochrones 
including overshooting and with $\eta$=0.2, extended along the TP-AGB
phase. At $Z$=0.0006 we display also values obtained with $\eta$=0.4
(dashed-dotted line -- see text for details).}
\label{SBFk}
\end{figure*}

\begin{figure*}
\epsscale{1}
\plotone{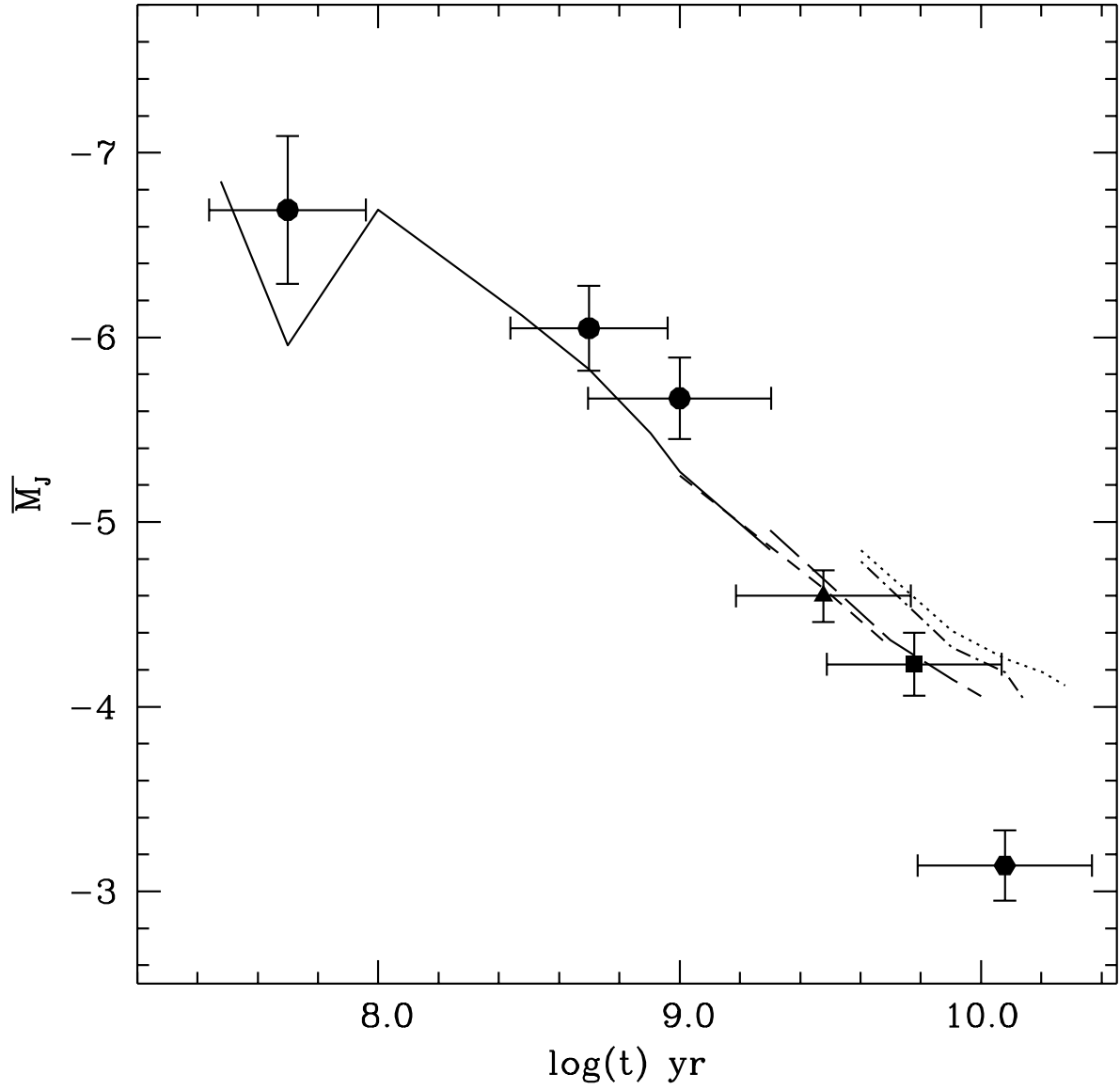}
\caption{As in Fig.~\ref{SBFk}, but for the $J$-band SBF magnitudes
(see text for details).}
\label{SBFj}
\end{figure*}

Figures~\ref{SBFk} and \ref{SBFj} display the observed SBF absolute magnitudes
in $K$ and $J$ (\cite{gonzalez_etal_04} 
assume $(m-M)_0=18.50\pm0.13$ for the Large Magellanic Cloud (LMC) and 
$(m-M)_0=18.99\pm0.05$ for the SMC) of 6 superclusters -- whose age range
is covered by our models -- as a function of age. 
Different symbols correspond to different mean 
metallicities; the three
youngest superclusters share all the same metallicity. Horizontal
error bars correspond to the age spread around the mean supercluster
ages, as given by \citet{gonzalez_etal_04}.

We have compared these data with the SBF magnitudes obtained from our
isochrones, for the appropriate mean metallicities of the individual superclusters. 
Notice that for each individual $Z$ considered, we have plotted the
theoretical SBF magnitudes only in the age range spanned by the one
(or more) supercluster with that mean value of $Z$. We employed the 
\textit{BaSTI} models with $\eta$=0.2, 
including overshooting in the appropriate age range. For the
lowest metallicity (oldest) supercluster we display also the results
obtained with $\eta$=0.4.
Theory is able to fit the data points within the observational
errors, the only exception being the $J$-band SBF of the 
oldest supercluster. 

\begin{figure*}
\epsscale{1}
\plotone{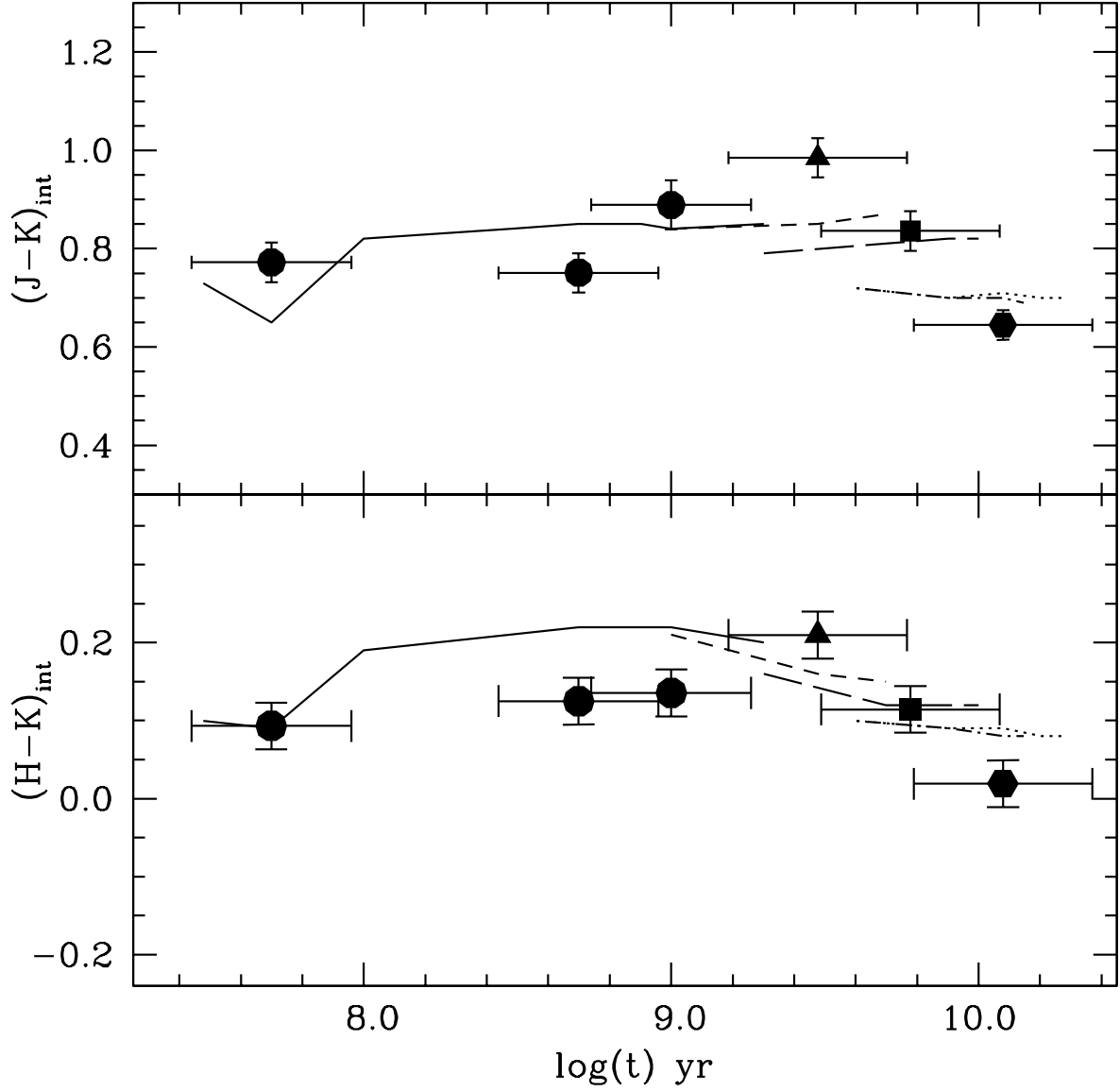}
\caption{As in Fig.~\ref{SBFk}, but for the integrated $(H-K)$ and
$(J-K)$ colors of Magellanic Cloud superclusters (see text for
details).}
\label{HKint}
\end{figure*}

\begin{figure*}
\epsscale{1}
\plotone{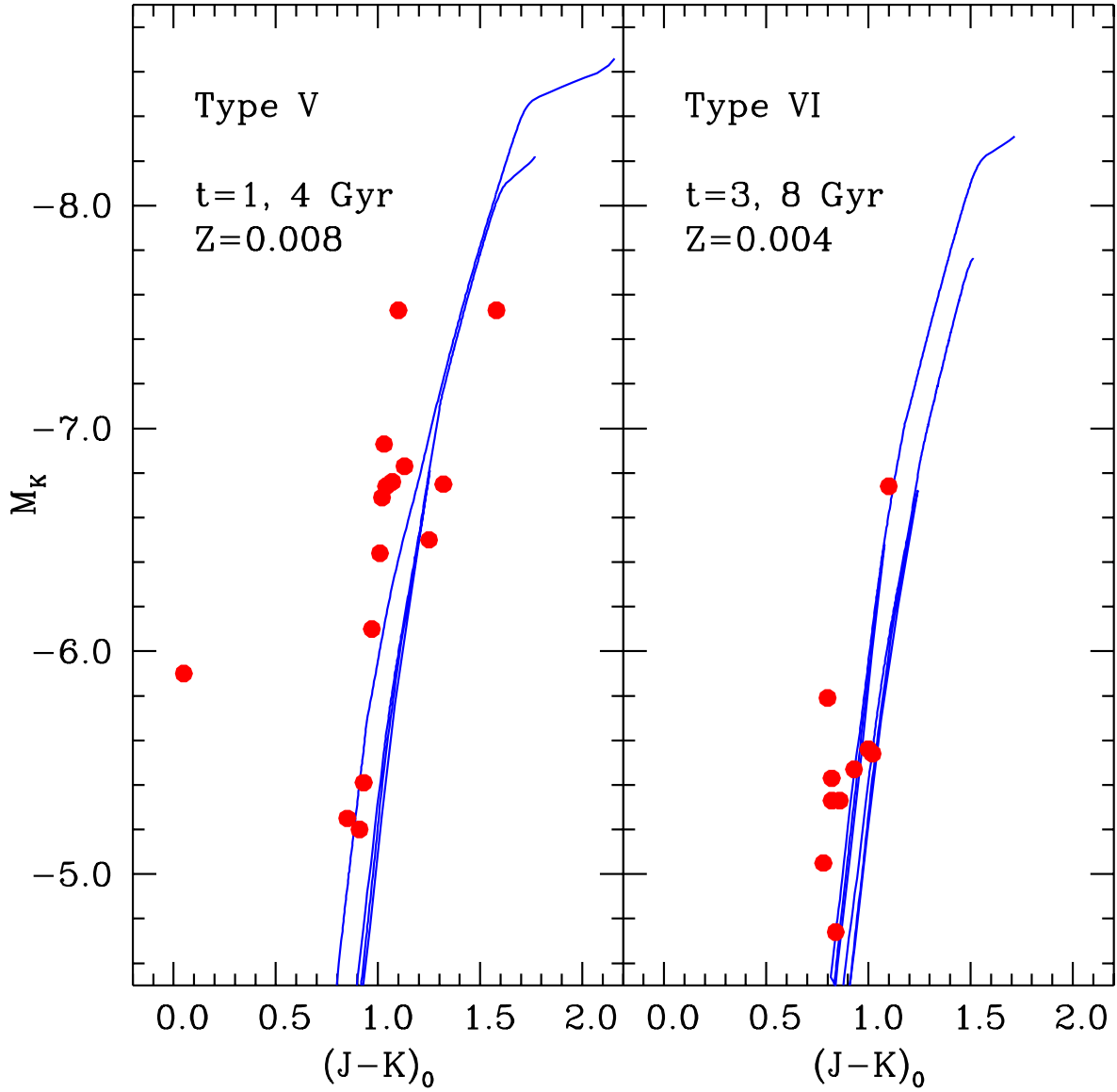}
\caption{Cumulative $K-(J-K)$ CMDs of AGB stars in a sample of LMC
clusters. Theoretical isochrones for the appropriate age and
metallicity range are also displayed (see text for details).}
\label{CMDcl}
\end{figure*}

Figure~\ref{HKint} compares the dereddened integrated $(J-K)$ and
$(H-K)$ colors provided by \cite{gonzalez_etal_04} for the 6 
superclusters -- transformed to the Johnson system as done for the SBF
-- with theoretical predictions from the same models employed for the SBF analysis. 
Theoretical colors appear consistent with the observational
counterpart; the mean difference between theory and observations is
equal to $\sim$0.03 mag in both $(J-K)$ and
$(H-K)$. This value can be considered a
reasonable estimate of the accuracy of our predicted integrated
near-IR colors.
Finally, Fig.~\ref{CMDcl} compares the combined $K-(J-K)$ CMDs of AGB stars in a
sample of LMC clusters, with theoretical isochrones. The individual
stars have been coadded according to the SWB class of the parent
clusters. Reddenings are from \citet{frogel_cohen_82} and we used an LMC
distance modulus $(m-M)_0$=18.50.
The metallicities and ages of the theoretical
isochrones have been selected on the basis of the SWB class following
again \citet{gonzalez_etal_04}. One can notice a good overlap between
the theoretical AGB sequences and the observed near-IR CMDs.

\section{Isochrone Maker and Track interpolation program} \label{iso_maker}

We have realized a dedicated WEB interface that allows a direct computation of isochrones for 
any given choice of age, for the chemical compositions and within the 
age ranges listed in tables~\ref{tab1} and \ref{tab2}.
It also enables the user to compute interpolated 
evolutionary tracks in the mass range $0.5 - 10M_\odot$ for any
chemical composition in the grid. A linear interpolation on a
point-by-point basis between two neighbouring evolutionary
tracks stored in the database is performed, to determine the track for the specified
mass. The mass spacing of the tracks included in \textit{BaSTI} is small
enough that a linear interpolation is sufficient to guarantee a high
accuracy. Once the track is computed, it is immediately transposed in
the chosen photometric system.

\begin{figure*}
\epsscale{0.8}
\plotone{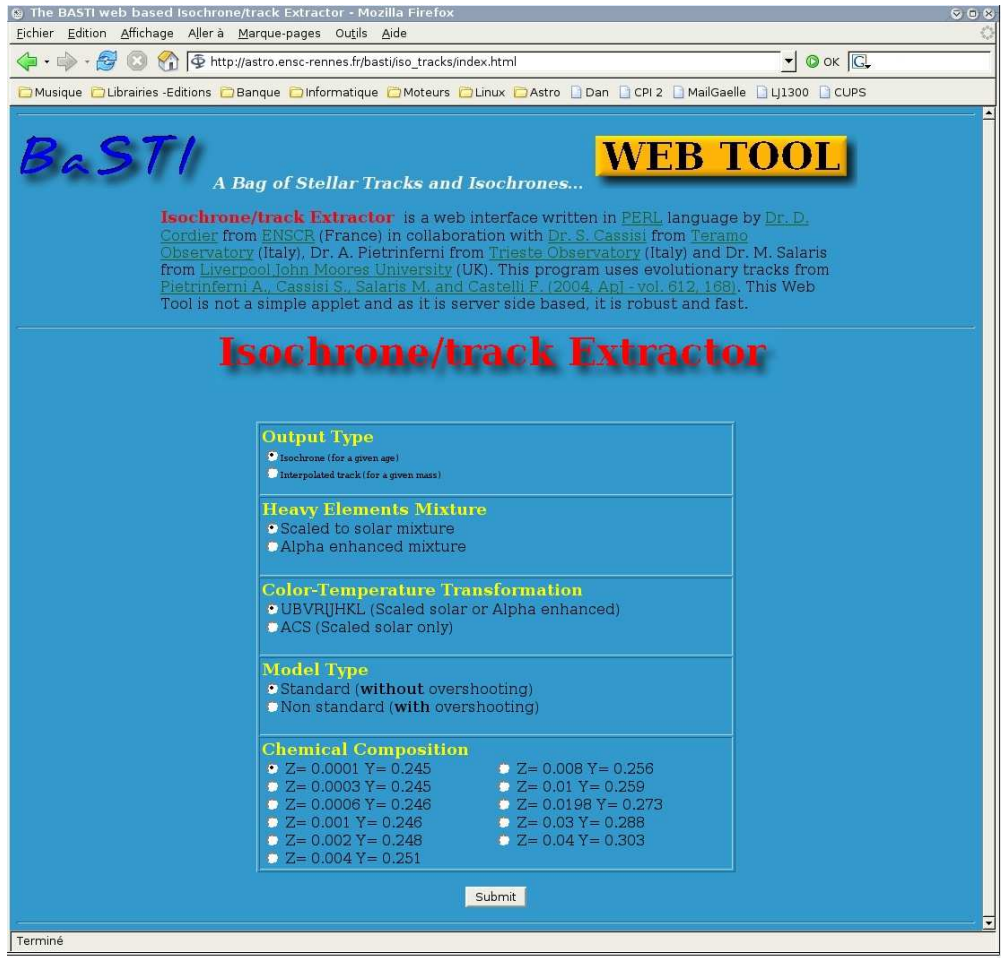}
\caption{Web interface for the tool dedicated to isochrones and
evolutionary tracks.}
\label{fig_iso_maker}
\end{figure*}

The use of this interface is particularly simple: first of all one has
to get to the server using this URL:
\begin{center}
       \verb+http://www.oa-teramo.inaf.it/BASTI+
\footnote{In case of Network and/or technical problem with this
official web site, the user can connect to the D. Cordier's personal site:
\texttt{http://astro.ensc-rennes.fr} where the \textit{BaSTI web-tools} have been mirrored
in {\it the BaSTI Web Tools} section.}
\end{center}
and click on the link \textit{Isochrones - Tracks} in the Web Tools
section; a page like the one displayed
in Fig.~\ref{fig_iso_maker} will appear. The user has now to choose
among a number of options:

\begin{itemize}

\item{the output type, i.e., isochrone or interpolated
evolutionary track;}

\item{the heavy elements mixture, i.e. scaled-solar or $\alpha-$enhanced;}

\item{the photometric system of interest; at the moment we allow to
choose between the standard $UBVRIJHKL$ filters and
the ACS-HST ones;}

\item{core convection treatment, i.e. with or without convective core overshoot;}

\item{the initial chemical composition ($Y, Z$).}
\end{itemize}

Once these input parameters are fixed, the next screen allows the
choice of the isochrone
age (in Myr) or the track initial total mass in solar units, and the 
job submission. Outputs (in the same format of the tracks and
isochrones stored in the database) are directly sent to the user's browser.
%-----------------------------------------------------
\section{Luminosity Function Program} \label{lum_func}

The isochrones stored in the \textit{BaSTI} database, or 
obtained with the \textit{web-tool} described in the previous section,  
can be used as input for the tool described here, that computes the 
luminosity functions (e.g. differential and cumulative
star-counts as a function of the magnitude in a given wavelength band)
of single-age, single-metallicity stellar populations. 

To run this \textit{web-tool} the user has to provide 4 parameters:
the number of isochrones whose luminosity functions are needed (maximum 20); 
the photometric filter (in the $UBVRIJKLH$ system); 
the Initial Mass Function (IMF) exponent;  
the size of the magnitude bins within which the star counts are
performed (e.g., 0.15~mag; these bins have to be
larger than $0.05$ mag). We adopt for the
IMF a form of this type: ${{dN_\star}\over{dM}}= C M^{-\alpha}$ where
${dN_\star}$ is the number of stars formed with mass between $M$ and
$M+dM$, and $C$ is a normalization constant whose value is fixed 
by imposing that the total number of stars populating the isochrones 
is equal to $10^6$. A value of $\alpha$ equal 
to 2.35 corresponds to the \citet{salpeter55} IMF.

One has then to choose between the options of computing the luminosity
functions only until the RGB tip, along the whole isochrone until the
Early-AGB, or including also the TP-AGB phase.
Star counts are then computed analytically by convolving
the IMF with the distribution of the initial stellar
mass values along the isochrones. For example, the number of stars $N$
between two consecutive points $i-1$ and $i$ along a chosen 
isochrone is given by:  
$N= {C\over{1-\alpha}}(M_{i}^{1-\alpha}-M_{i-1}^{1-\alpha})$. 

As a final step one has to upload the isochrone files, and submit the
job. Results are directly sent to the user
Internet browser as ASCII data, that are divided
into several sections, each of them corresponding to one individual 
isochrones. The input  parameters are recalled: 
metallicity, IMF exponent, age, etc. The data themselves are displayed
in three columns: the mean magnitude of the
bin, the differential ($\log N$ -- where $N$ denotes the star counts per magnitude bin
as a function of the magnitude) and cumulative ($\log N+$ -- where
$N+$ denotes the sum of the star counts from the faintest bin to the
actual one, as a function of their magnitude) luminosity functions.

%---------------------------------------------------------------  
\section{Synthetic CMD code} \label{pop_synth}

The computation of synthetic CMDs is often required to interpret 
observations of resolved or unresolved stellar
populations. The simplest form of a synthetic CMD is an isochrone,
that represents the sequence occupied by stars all formed at
the same time and with the same initial chemical composition. Clearly
an isochrone does not directly contain information about the number of
stars populating the various CMD branches, nor the effect of
photometric errors and reddening, and cannot represent composite
stellar populations made of multiple generations of stars. 
To include these effects one needs a synthetic CMD generator. In
the \textit{BaSTI} website the user can access the SYNTHETIC MAN(ager) code 
that is an evolution of the version 
briefly described in \citet{pietrin_etal_04}. This code 
computes magnitudes and colors of objects belonging to a synthetic
stellar population with an arbitrary Star Formation History (SFH). 
The program employs a grid of isochrones with ages between 30~Myr and 
14~Gyr (we employ 49 isochrones extended until the end of 
the TP-AGB phase, for each of the 11 metallicities included in our model grid). 

The SYNTHETIC MAN code is based on a Monte-Carlo algorithm,  
that allows one to include in a simple way 
observation-related effects like reddening/distance/metallicity 
dispersion and photometric errors, and to identify on a
star-by-star basis the various types of pulsating stars that 
may be present in the synthetic population.
Synthetic CMDs and observation-related effects 
can also be efficiently determined
by means of analytical integrations, as done in \citep{dolphin_02}.
It is worth noticing 
that 
\citet{skillman_etal_03}
 applied three different methods
to determine the SFH of the galaxy IC1613. One of them is the analytical
technique by
\citet{dolphin_02},
another one is a Monte-Carlo method 
similar to the one we implemented. A comparison of the results
with the different techniques shows striking agreement.

The general structure of the SYNTHETIC MAN code is as follows.
An SFH file has to be specified first. The SFH contains a grid of $N$ ages $t_i$ 
(the upper limit for $N$ is set to 200) increasing with 
increasing running index $i$   
(the present time is assigned $t$=0) and for each age one has to
specify a scale factor $SF_{i}$ proportional to the
relative number of stars formed at that time. Also,  
the metallicity (parametrized by [Fe/H]) of the stars formed at that
age has to be given (${\rm [Fe/H]}_i$) together with a 1$\sigma$ Gaussian spread around this value.
The scale factor and [Fe/H] for ages between $t_i \le {t} < t_{(i+1)}$ are then assumed to be equal to 
$SF_i$ and ${\rm [Fe/H]}_i$, respectively. In other words, the SFH is considered to 
to be a sequence of step functions. 
The scale factor and [Fe/H] at $i=N$ are not 
considered and can be set to any arbitrary value; the last point in the SFH file is important only 
because it provides the upper age-limit for the simulated stellar population. 
If one or more single burst stellar populations are needed, one or more pairs 
of identical ages (i.e. $t_i=t_{(i+1)}$) need to be included in the SFH file.  
 
After the SFH is read by the code, the following cycle starts, running 
along the SFH index, from $i=0$ to $i$=$N$.   
The age difference between two generic $t_i$ and 
$t_{(i+1)}$ is computed, and for each star formed in this 
time interval (the number of stars 
formed in a generic time interval is determined from 
the input SFH) a random value of the stellar age 
$t_i \leq t < t_(i+1)$ is drawn with a flat probability distribution, 
together with a mass $M$ selected randomly according to a user specified IMF 
(the default mass range goes from 0.1$M_{\odot}$ to 120$M_{\odot}$, although a different lower mass 
limit can be specified by the user)  
and a random value of [Fe/H] assigned according to the value of ${\rm [Fe/H]}_i$ plus 
the 1$\sigma$ spread given in the SFH. 
With the three specified values of $t$, $M$ and [Fe/H] the program interpolates quadratically in 
age, metallicity and then mass among 
the isochrones in the grid, to determine the star luminosity, effective
temperature, actual value of the mass (in principle different from the
initial value because of mass loss) and photometric properties in the $UBVRIJHKL$ system. 

For each generated mass an additional random number determines whether the star is member
of an unresolved binary system (the percentage of unresolved binaries
has to be specified as input); if this is true, the mass of the second
component is selected randomly following \citet{woo_etal_2003} and
the fluxes of the two unresolved components are properly added.

Once the photometric properties of an object (single or belonging to
an unresolved binary system) are determined, depth effects of the
synthetic stellar population are simulated probabilistically 
according to a uniform stellar distribution with a user specified total depth (in mag). 
After depth effects are included, the individual magnitudes are
modified by adding the effect of extinction (another input parameter; we have
employed the extinction ratios from the Asiago Database on Photometric Systems --see 
\citet{moro_munari_2000}-- and $R_V$=3.1) 
and then further perturbed according to a Gaussian distribution, to simulate the
photometric errors, with a user specified 1$\sigma$ width.
The star [Fe/H] value is also perturbed according to a
Gaussian error with user specified 1$\sigma$ width, to mimic the
effect of spectroscopic observational errors. 
As an additional option, the code can search for variable stars, 
according to their location with respect to the boundary of RR Lyrae
and Cepheid instability strips; pulsation periods are then determined.  
Table~\ref{tbl-variables} shows literature references for the relevant
type of variables.

Once all stars formed between 
ages $t_i$ and $t_{(i+1)}$ are generated, 
the next interval ($t_{(i+1)} \le t < t_{(i+2)}$) is 
considered, and the cicle continues, ending when all stars 
in the final age bin $t_{(N-1)} \le t < t_{(N)}$ are generated. 

As a technical comment, 
if the value of $M$ for a star generated by the Monte-Carlo procedure 
is too large to be still evolving at its age $t$, or 
is lower than 0.5$M_{\odot}$, the contribution to the total mass of 
stars formed in the population is taken into account, but the star photometric 
properties are not determined. 
The total mass of the objects formed, the total number of the stars whose photometric properties 
have been determined (in principle different from the total number generated for 
the reasons mentioned above) plus integrated magnitudes (in the $UBVRIJHKL$ system)  
of the synthetic population are also computed.

Like all Monte-Carlo based simulations, our program makes an extensive
use of random numbers. Our random number generator (initially written
by \cite{james90}) needs \textit{seeds}
to be initialized; these seeds can be either provided by the user or got
from \texttt{http://www.random.org} which derived \textit{true} random
numbers from atmospheric electromagnetic noise.

The web interface to SYNTHETIC MAN can been found in the 
\textit{BaSTI} section called: \textit{Synthetic Color - Magnitude Diagrams}. This software
is as simple to use as the others but a registration is required
before its first use. The user receives a notification by e-mail
when the computations are completed. \textbf{In order to get an user
identification, one needs to 
contact S. Cassisi\footnote{\email{cassisi@oa-teramo.inaf.it}}
or D. Cordier\footnote{\email{daniel.cordier@ensc-rennes.fr}}}.
In the following we summarize the input parameters for a run of
SYNTHETIC MAN.

\begin{itemize}

\item{A scale factor for the SFH. As explained before, the SFH contains the
relative weight of the star formation episodes in the various age bins. In
order to fix the absolute number of objects in each stellar
generation, a scale factor has to be chosen. The product of this scale
factor times the relative weight in the input SFH gives the total
number of stars created in a generic age bin, with mass between 0.1 (or a user specified 
lower mass limit) and 120$M_{\odot}$.
If more than $2 \ 10^6$  objects are expected to be formed at a given
time step, the program will stop to highlight the excessive amount of
computational time needed for the whole simulation.}

\item{A choice for the mean photometric error. The user can adopt 
a 1$\sigma$ constant error that it is applied to all stars in the simulation, in 
all 9 photometric bands. The size  
of this constant error (between 0.0 and 1.0~mag) has to be specified. Alternatively, the user
can choice to adopt an error varying with the actual 
star magnitude and/or photometric band. The exact values have to be specified 
by the user in an appropriate input file. This photometric error input file contains 
18 columns and an arbitrary number $N$ of rows (upper limit $N$=200). 
The first 9 columns display $N$ magnitudes $M_{\lambda}^{i}$ 
($i$ running from 1 to $N$) in order of increasing values, 
for the $UBVRIJHKL$ photometric bands respectively.
The remaining 9 rows are the corresponding 1$\sigma$ photometric errors 
in the $UBVRIJHKL$ filters, respectively (e.g., 
$\sigma_U$, $\sigma_B$, $\sigma_V$ and so on). The choice of the error 
for any individual synthetic star proceeds as follows. If $M_V^{\star}$ is the star  
magnitude in the $V$ band, the program searches 
for a pair of neighboring tabulated $M_V^i$ magnitudes such that 
$M_V^i < M_V^{\star} \leq M_V^{i+1}$, and assigns to the star 
$\sigma_V$ the value tabulated for $M_V^{i+1}$.
If $M_V^{\star}$ is smaller (larger) than $M_V^1$ ($M_V^N$) the error 
corresponding to $M_V^1$ ($M_V^N$) is used.
The same procedure is repeated for all other photometric 
bands considered.} 

\item{The 1$\sigma$ mean
spectroscopic error (between 0.01 and 1.0~dex).} 

\item{A choice for range considered in the stellar mass random extraction. 
An upper limit of 120$M_{\odot}$ is always considered. The user can decide to use a different lower mass limit 
(larger than 0.1$M_{\odot}$) that has then to be specified by the user.} 

\item{The total spatial
depth of the population (between 0.0 and 10.0~mag).}

\item{The colour excess
$E(B-V)$ (between 0.0 and 10.0~mag).}

\item{The fraction of
unresolved binaries.} 

\item{The minimum mass ratio for the unresolved binary
systems, that enters the \citet{woo_etal_2003} relationship (between
0.0 and 1.0)} 

\item{The IMF. The user can adopt a power law of the form $M^{-x}$ is employed. 
The exponent $x$ has then to be specified. Alternatively, the user is allowed to
choice the \citet{kroupa93} IMF.}

\item{Choice of the isochrone set to use according to the value of an integer index 
between 1 and 8. The correspondence between integer value and isochrone 
set is reported in Table~\ref{tbl-isoset}.} 

\item{Possibility to search for pulsating variable stars harbored by
the synthetic population, and compute their periods. An integer value equal to 1 enables the search, 
a value equal to 0 prevents the search.}

\item{Input the desired SFH. A number of pre-specified SFHs for selected galaxies are also
provided (see Table~\ref{tbl-sfh}).}

\item{As all simulations based on a Monte-Carlo method, our program needs \textit{seeds} for
        random numbers generator initialization. The user can choose
        the values of the seeds,  or let
	the program automatically get seeds from the \textit{server of entropy} 
	\texttt{http://www.random.org}. We underline that all computations initialized with
	the same seeds will lead to identical outputs.}
\end{itemize}

After specifying the input parameters, users can start their
computations by  clicking the ``\textit{Submit}'' button. The calculation 
may take some time (from a few minutes to
hours, mainly depending on the requested number of stars). After the 
calculation is completed the user receives an e-mail
as notification. As the size of the output files can amount to several MBytes, 
they cannot be sent by e-mail and a solution involving a 
dynamically generated web page has been found preferable. In the 
output web-page four data files 
are available: 
\begin{itemize}
  \item \verb+Synth_Pop_*_user.in+: this ASCII file recalls the parameters used for
        the calculation.
  \item \verb+Synth_Pop_*_.HRD.gz+: this gzipped file contains data 
        for the individual stars, in which one can find the following quantities: 
        identification number, log($t$) (yr), [Fe/H], the value of the actual
        mass, ${\rm \log(L/L_{\odot})}$, log(${\rm T_{eff}}$), $M_V$, $(B-V)$, $(U-B)$,
        $(V-I)$, $(V-R$), $(V-J$), $(V-K$), $(V-L)$, $(H-K)$, the value of the initial
        mass, the initial mass of the unresolved binary companion (if present
        -- for stars without companion this quantity is set to 0.0), logarithm 
        of the pulsation period (in days -- if the search for variable is off
        or if the star is not pulsating this quantity is set to
        99.99), an index denoting the type of variable (see Tab.~\ref{tbl-variables} for explanations).
        This file can be uncompressed with the standard GNU software
        \verb+gunzip+. As this file can be relatively large,
        the downloading is automatically forced when users click on the link. 
  \item \verb+Synth_Pop_*_.sfh+: contains the Star Formation History data used for the calculations.

  \item \verb+Synth_Pop_*_.INT_PROPERTIES+: file with the integrated $UBVRIJHKL$ absolute
         magnitudes produced by the stars evolving in the synthetic 
         population, several selected integrated colors, the total number of
         stars with computed photometric properties, the total mass of stars formed (within the mass range 
         specified in the input file) according to the specified SFH 
         and a summary of the various 
         types of variable stars found (if the search for variables is off the values are all equal to zero).
\end{itemize}

The symbol * denotes a string chosen by the WEB tool, to
avoid that output files corresponding to a specific simulation 
are overwritten by those related to a different run.
The string is selected according to the time when 
the numerical run has been launched. For example, if the run 
is launched at 16.51.34 of May the 3rd of the year 2006, the
string will be \lq\verb+May_3_16.51.34_2006+\rq

\begin{figure*}
\epsscale{1.0}
\plotone{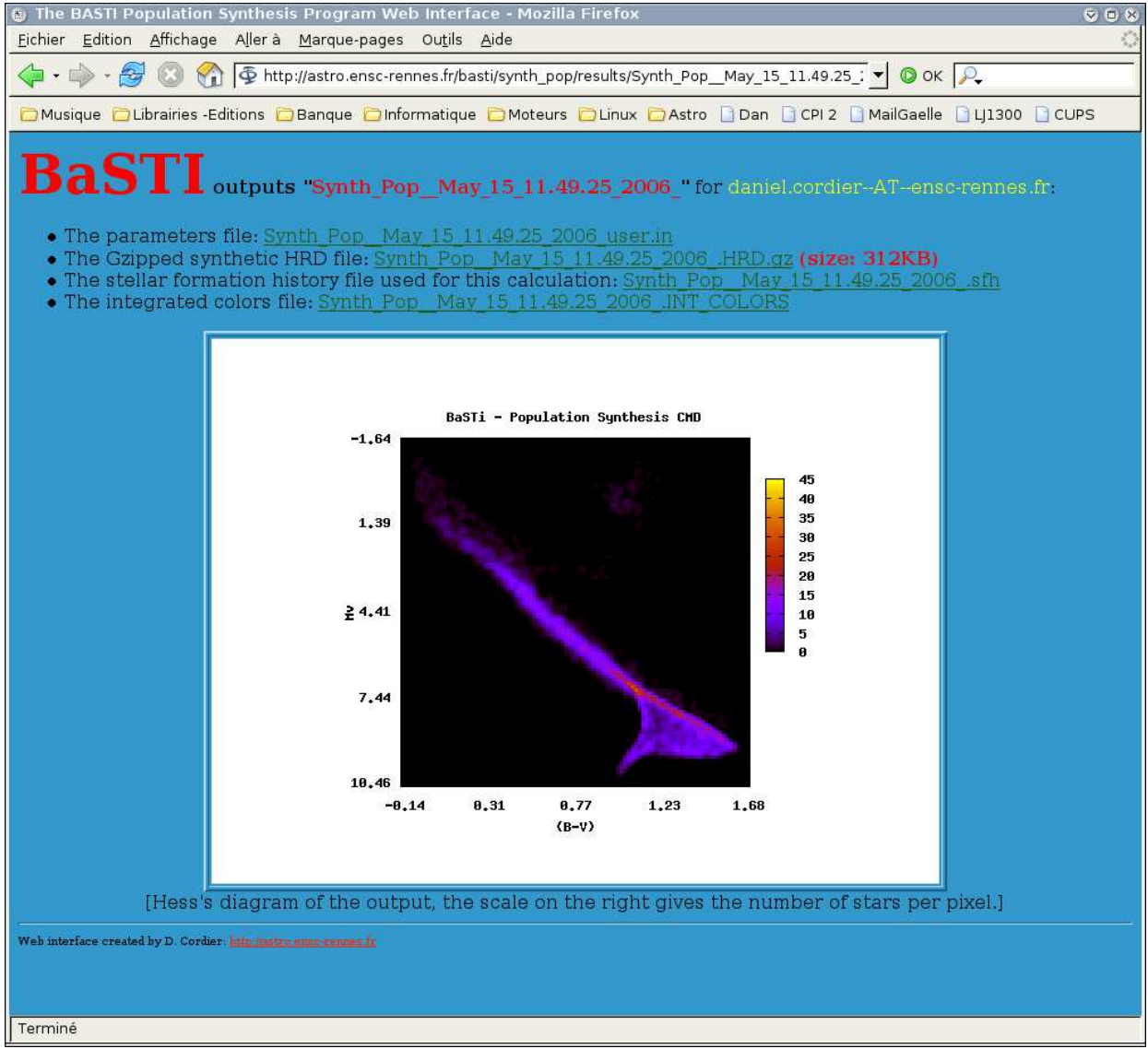}
\caption{Example of the output web-page for the SYNTHETIC MAN \textit{web-tool}.}
\label{fig_synthman}
\end{figure*}

An example of the output web-page is displayed in
Fig.~\ref{fig_synthman}. The $V-(B-V)$ Hess's diagram taking into account all
the synthetic objects drawn for the simulation is also automatically displayed, stars numbers being
encoded with colors.
 
The user should be aware that -- for obvious reasons -- web pages displaying program outputs are
deleted by the software manager one month after the end of the computation.

%---------------------------------
\section{Conclusion} \label{concl}

%% Included in this acknowledgments section are examples of the
%% AASTeX hypertext markup commands. Use \url without the optional [HREF]
%% argument when you want to print the url directly in the text. Otherwise,
%% use either \url or \anchor, with the HREF as the first argument and the
%% text to be printed in the second.

In this paper we have presented the content of the \textit{BaSTI}
database for stellar population synthesis studies. In particular, we
have discussed the recent extension of the models until the end of the
TP-AGB phase, and described three \textit{web-tools} to compute
synthetic CMDs, luminosity functions and user-specified isochrones and
stellar evolution tracks. 

In the near future the \textit{BaSTI} database will be developed following
two directions: the extension of the database itself and the development of
an \textit{on-line based} version of our stellar evolution code.\\
We plan to extend the mass range spanned by our database, by adding 
tracks for very low mass objects (below 0.5~$M_{\odot}$) and
very massive stars (above 10~$M_{\odot}$), the latter until C-burning ignition. 
%The grid of models being also extended to cover
%the full asymptotic giant
%branch domain with the aim of a full coverage of thermal pulses
%phase.
\\
With a \textit{Web-based} version of our evolution code, people who are not
expert in stellar evolution calculations will be able to compute models with their
customized set of parameters (mixing length value, extension of the overshooting region, metallicity, etc),
avoiding inaccurate interpolation and/or extrapolation between and/or from existing models.
That could also allow for more complex tuning like: opacity tables or nuclear rate table
switching, change of equation of state, etc. We think that this \textit{Web-based} evolutionary code, as well
as the \textit{BaSTI} database could be relevant tools for the whole scientific community within the
ongoing \textit{Virtual Observatory}'' project \citep{pasian04}.

\acknowledgments

We wish to thank an anonymous referee for suggestions that helped us to improve the content
and readability of the manuscript.
D. Cordier thanks the Ecole Nationale Sup\'erieure de Chimie de Rennes for working facilities and M.J. Goupil
and Y. Lebreton for so many things. We wish also to acknowledge all people
who have already used and/or will be using the 
\textit{BaSTI web-tools} for their own research, and 
who will send us their comments/suggestions. We warmly thank E. Sandquist for all the useful
comments and for pointing out the existence of some problems in a preliminary
version of our WEB tools. S.C. warmly thanks financial support from MIUR and INAF.

\appendix

\section{BaSTI Web Tools: some technical aspects}

\begin{figure*}
\epsscale{0.6}
\plotone{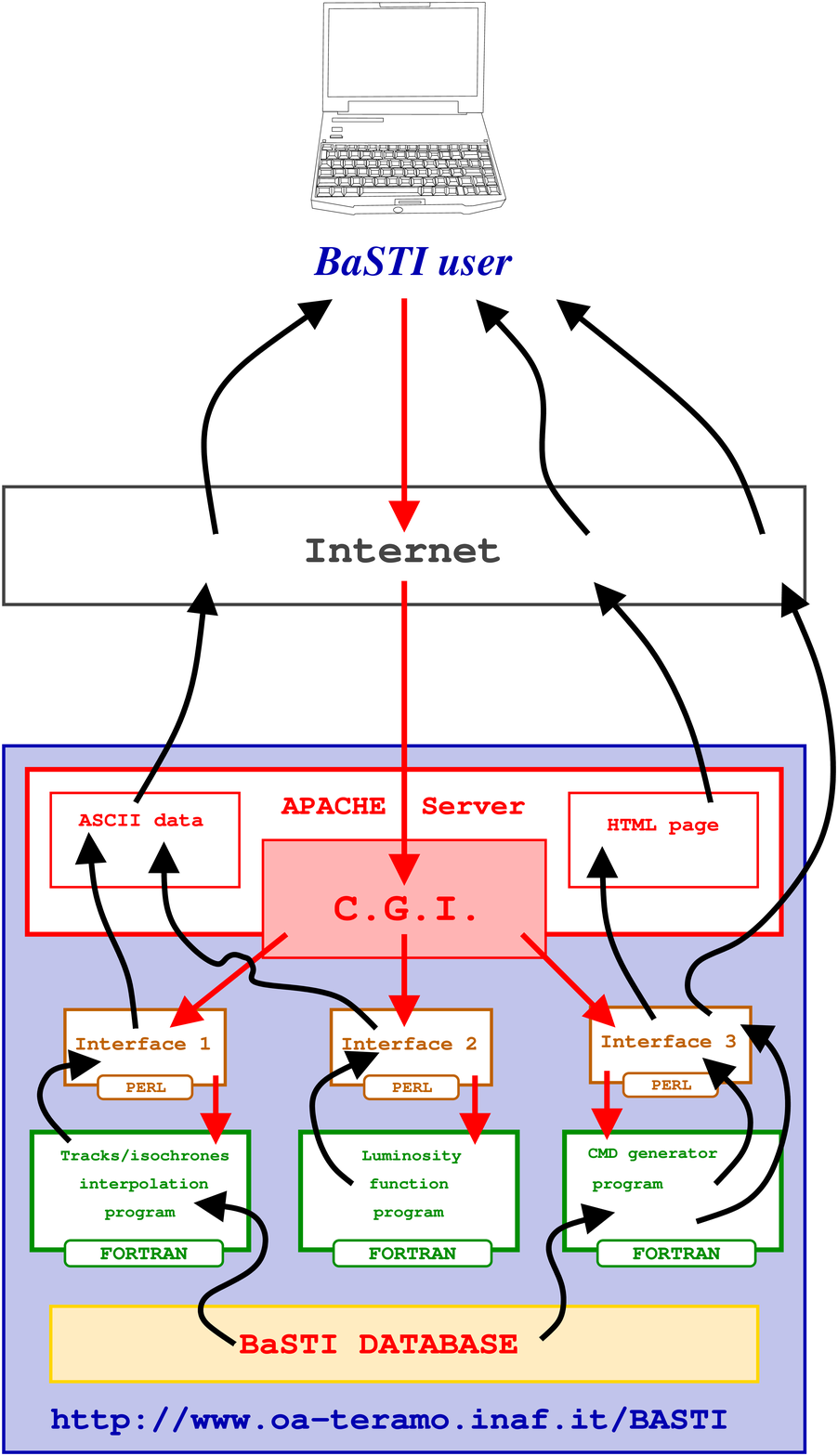}
\caption{Data fluxes in the \textit {BaSTI web tools}. Random number generator seeds provided by
\texttt{http://www.random.org} has been omitted.}
\label{datafluxes}
\end{figure*}

The World Wide Web (WWW), originally conceived for document delivery, has evolved into a 
medium supporting interactive computations. One way for a web server (in our case
\texttt{APACHE}, see \texttt{http://www.apache.org}) to interact with data-generating
programs is the well known Common Gateway Interface (hereafter CGI). Input parameters and data
provided by the user are passed to our three FORTRAN programs through PERL programs.
PERL is probably the most common CGI scripting language, for which an huge collection of modules
is available in the Comprehensive Perl Archive Network 
(see \texttt{http://www.cpan.org}).\\
The tracks/isochrones interpolation program and the luminosity function program
are simply driven with a specific PERL interface, their outputs are
directly displayed as an ASCII flux which can be saved in the user's hard disk and used by other
programs without any difficulty.\\ 
The CMD generator is probably -- at the moment-- our more sophisticated WWW-based software.
When it is invoked, a computation identification string of characters (CISC) is assigned.
The files are handled and stored  with this CISC as part of the file name --so even if multiple
users from different parts of the world submit simulations at the same time, they can all be processed
separately. When the computation is completed, an HTML page is created on-the-fly, displaying
data output files and a pre-visualization made with \texttt{GNUplot}. The Perl module
\texttt{Net} allows automatic e-mail sending for user notification.\\
Figure~\ref{datafluxes} summarizes the \textit{BaSTI Web-Tools} general
scheme. Some details have been omitted, like interaction with the plotting sofware
\texttt{GNUplot} and random seeds download from \texttt{http://www.random.org}.
At the moment all calculations are treated after each user's clicking,
and consequently several
computations can be running at the same time on our machine, an
occurrence that could seriously affect 
the computation time. Depending on the success of our on-line programs, we could 
implement as a new feature a more elaborated computations managing system that could
"queue" the users requests, decreasing individual computation times.


%%%%%%%%%%%%%%%%%%%%%%
% Bibliography
\bibliographystyle{apj} % style apj.bst
\bibliography{./apj-jour-perso,./bibliographie_MASTER_BASTI}
%\input{ms.bbl}

%%%%%%%%%%%%%%%%%%%%%%%%%%%%%%%%%%%%%%%%%%%%%%%%%%%%%%%%%%%%%%%%%%%%%%%%%%%%%%%%
\clearpage

\begin{table}
\begin{center}
\caption{Initial chemical composition of the model grid.}
\begin{tabular}{cccc}
\tableline
$Z$ &  $Y$  &  [Fe/H] (scaled solar mixture) & [Fe/H]
($\alpha$-enhanced mixture) \\
\tableline
                   0.0001 & 0.245 & $-$2.27 & $-$2.62 \\        
                   0.0003 & 0.245 & $-$1.79 & $-$2.14 \\ 
                   0.0006 & 0.246 & $-$1.49 & $-$1.84 \\ 
                   0.0010 & 0.246 & $-$1.27 & $-$1.62 \\                       
                   0.0020 & 0.248 & $-$0.96 & $-$1.31 \\ 
                   0.0040 & 0.251 & $-$0.66 & $-$1.01 \\ 
                   0.0080 & 0.256 & $-$0.35 & $-$0.70 \\ 
                   0.0100 & 0.259 & $-$0.25 & $-$0.60 \\ 
                   0.0198 & 0.273 &  +0.06  & $-$0.29 \\ 
                   0.0300 & 0.288 &  +0.26  & $-$0.09 \\ 
                   0.0400 & 0.303 &  +0.40  &  +0.05  \\ 
\tableline
\end{tabular}
\end{center}
\label{tab1}
\end{table}

%%%%%%%%%%%%%%%%%%%%%%%%%%%%%%%%%%%%%%%%%%%%%%%%%%%%%%%%%%%%%%%%%%%%%%%%%%%%%%%%
%%%%%%%%%%%%%%%%%%%%  TABLES
%%%%%%%%%%%%%%%%%%%%%%%%%%%%%%%%%%%%%%%%%%%%%%%%%%%%%%%%%%%%%%%%%%%%%%%%%%%%%%%%
\clearpage

\begin{table*} 
\centering 
\caption{Summary of the \textit{BaSTI} model and isochrone database.}
\begin{tabular}{|c|c|c|c|c|c|c|c|c|}

\hline 
\multicolumn{1}{|c|}{ {mixture}} & 
\multicolumn{4}{|c|}{ {scaled-solar}} & \multicolumn{4}{|c|}{ 
{$\alpha$-enhanced}} \\ 
 
\hline 
\multicolumn{1}{|c|}{ {$\eta$}} & 
\multicolumn{2}{|c|}{ {0.2}} & \multicolumn{2}{|c|}{ {0.4}} &  
\multicolumn{2}{|c|}{ {0.2}} & \multicolumn{2}{|c|}{ {0.4}}  \\ 
 
\hline 
\multicolumn{1}{|c|}{{$\lambda_{OV}$}} & 
\multicolumn{1}{|c|}{ {0}} & \multicolumn{1}{|c|}{ {0.2}} &  
\multicolumn{1}{|c|}{ {0}} & \multicolumn{1}{|c|}{ {0.2}} &  
\multicolumn{1}{|c|}{ {0}} & \multicolumn{1}{|c|}{ {0.2}} &  
\multicolumn{1}{|c|}{ {0}} & \multicolumn{1}{|c|}{ {0.2}} \\ 
 
\hline 
 
\multicolumn{1}{|c|}{{$N^O$} tracks} & 
\multicolumn{1}{|c|}{ {20}} & \multicolumn{1}{|c|}{ {20}} &  
\multicolumn{1}{|c|}{ {40}} & \multicolumn{1}{|c|}{ {20}} &  
\multicolumn{1}{|c|}{ {20}} & \multicolumn{1}{|c|}{ {20}} &  
\multicolumn{1}{|c|}{ {40}} & \multicolumn{1}{|c|}{ {20}} \\ 
 
\hline 
 
\multicolumn{1}{|c|}{{$M_{min}$}(M$_\odot$)} & 
\multicolumn{1}{|c|}{ {0.5}} & \multicolumn{1}{|c|}{ {1.1}} &  
\multicolumn{1}{|c|}{ {0.5}} & \multicolumn{1}{|c|}{ {1.1}} &  
\multicolumn{1}{|c|}{ {0.5}} & \multicolumn{1}{|c|}{ {1.1}} &  
\multicolumn{1}{|c|}{ {0.5}} & \multicolumn{1}{|c|}{ {1.1}} \\ 
 
\hline 
 
\multicolumn{1}{|c|}{{$M_{max}$}(M$_\odot$)} & 
\multicolumn{1}{|c|}{ {2.4}} & \multicolumn{1}{|c|}{ {2.4}} &  
\multicolumn{1}{|c|}{ {10}} & \multicolumn{1}{|c|}{ {10}} &  
\multicolumn{1}{|c|}{ {2.4}} & \multicolumn{1}{|c|}{ {2.4}} &  
\multicolumn{1}{|c|}{ {10}} & \multicolumn{1}{|c|}{ {10}} \\ 
 
\hline 
 
\multicolumn{1}{|c|}{{$N^O$} isoc.} & 
\multicolumn{1}{|c|}{ {63}} & \multicolumn{1}{|c|}{ {44}} &  
\multicolumn{1}{|c|}{ {54}} & \multicolumn{1}{|c|}{ {44}} &  
\multicolumn{1}{|c|}{ {63}} & \multicolumn{1}{|c|}{ {44}} &  
\multicolumn{1}{|c|}{ {54}} & \multicolumn{1}{|c|}{ {44}} \\ 
 
\hline 
 
\multicolumn{1}{|c|}{{$Age_{min}$}(Myr)} & 
\multicolumn{1}{|c|}{ {30}} & \multicolumn{1}{|c|}{ {30}} &  
\multicolumn{1}{|c|}{ {30}} & \multicolumn{1}{|c|}{ {30}} &  
\multicolumn{1}{|c|}{ {30}} & \multicolumn{1}{|c|}{ {30}} &  
\multicolumn{1}{|c|}{ {30}} & \multicolumn{1}{|c|}{ {30}} \\

\hline 
 
\multicolumn{1}{|c|}{{$Age_{max}$}(Gyr)} & 
\multicolumn{1}{|c|}{ {19}} & \multicolumn{1}{|c|}{ {9.5}} &  
\multicolumn{1}{|c|}{ {14.5}} & \multicolumn{1}{|c|}{ {9.5}} &  
\multicolumn{1}{|c|}{ {19}} & \multicolumn{1}{|c|}{ {9.5}} &  
\multicolumn{1}{|c|}{ {14.5}} & \multicolumn{1}{|c|}{ {9.5}} \\ 
 
\hline 
\multicolumn{1}{|c|}{{Color-$T_{eff}$}} & 
\multicolumn{8}{|c|}{{$UBVRIJHKL$ - ACS $HST$}} \\ 
\hline 
 
\end{tabular} 
\label{tab2} 
\end{table*} 
% 
% 
%%%%%%%%%%%%%%%%%%%%%%%%%%%%%%%%%%%%%%%%%%%%%%%%%%%%%%%%%%%%%%%%%%%%%%%%%%%%%%%%
\clearpage

\begin{table}
\footnotesize
\begin{center}
\caption{Variable type index and sources used in the population synthesis program.\label{tbl-variables}}
\begin{tabular}{cll}
\tableline
Variables Type Index & Variable Type                     & Source for Instability Strip Boundaries \\
\tableline
                   0 & no variable                       &  \\
                   1 & fundamental RR Lyrae              &  \citet{marconi_etal_03,di_Criscienzo_etal_04}\\
                   2 & first overtone RR Lyrae           &  \citet{marconi_etal_03,di_Criscienzo_etal_04}\\
                   3 & fundamental anomalous Cepheid     &  \citet{marconi_etal_04}\\
                   4 & first overtone anomalous Cepheid  &  \citet{marconi_etal_04}\\
                   5 & fundamental classical Cepheid     &  \citet{bono_etal_00b}\\
\tableline
\end{tabular}
\end{center}
\label{tab3} 
\end{table}

%%%%%%%%%%%%%%%%%%%%%%%%%%%%%%%%%%%%%%%%%%%%%%%%%%%%%%%%%%%%%%%%%%%%%%%%%%%%%%%%
\clearpage

\begin{table}
\footnotesize
\begin{center}
\caption{Index for the isochrone sets employed by the SYNTHETIC MAN code.\label{tbl-isoset}}
\begin{tabular}{cl}
\tableline
  Index & Isochrone set                    \\
\tableline
                   1 & Scaled solar, no overshooting, $\eta$=0.2   \\
                   2 & Scaled solar, overshooting, $\eta$=0.2      \\
                   3 & Scaled solar, no overshooting, $\eta$=0.4   \\
                   4 & Scaled solar, overshooting, $\eta$=0.4      \\
                   5 & $\alpha$-enhanced, no overshooting, $\eta$=0.2   \\
                   6 & $\alpha$-enhanced, overshooting, $\eta$=0.2      \\
                   7 & $\alpha$-enhanced, no overshooting, $\eta$=0.4   \\
                   8 & $\alpha$-enhanced, overshooting, $\eta$=0.4      \\
\tableline
\end{tabular}
\end{center}
\label{tab4} 
\end{table}

%%%%%%%%%%%%%%%%%%%%%%%%%%%%%%%%%%%%%%%%%%%%%%%%%%%%%%%%%%%%%%%%%%%%%%%%%%%%%%%%
\clearpage
\begin{table}
\begin{center}
\caption{Available Star Formation Histories.\label{tbl-sfh}}
\begin{tabular}{ll}
\tableline
Stellar Population & Source \\
\tableline
NGC6822   (global SFH)     & \citet{gallart_etal_96}\\
SMC       (global SFH)     & \citet{harris_zaritsky_04}\\
LMC (bar field)            & \citet{holtzman_etal_99}\\
Local disk                 & \citet{rocha_etal_00_II}\\
Milky Way bulge            & \citet{molla_etal_00}\\
Sextans A                  & \citet{dolphin_etal_03}\\
LGS3                       & \citet{miller_etal_01}\\
\tableline
\end{tabular}
\end{center}
\label{tab5} 
\end{table}

%%%%%%%%%%%%%%%%%%%%%%%%%%%%%%%%%%%%%%%%%%%%%%%%%%%%%%%%%%%%%%%%%%%%%%%%%%%%%%%%
%%%%%%%%%%%%%%%%%%%%  FIGURES
%%%%%%%%%%%%%%%%%%%%%%%%%%%%%%%%%%%%%%%%%%%%%%%%%%%%%%%%%%%%%%%%%%%%%%%%%%%%%%%%

\clearpage

%%%%%%%%%%%%%%%%%%%%%%%%%%%%%%%%%%%%%%%%%%%%%%%%%%%%%%%%%%%%%%%%%%%%%%%%%%%%%%%%

\clearpage

%%%%%%%%%%%%%%%%%%%%%%%%%%%%%%%%%%%%%%%%%%%%%%%%%%%%%%%%%%%%%%%%%%%%%%%%%%%%%%%%

\clearpage

%%%%%%%%%%%%%%%%%%%%%%%%%%%%%%%%%%%%%%%%%%%%%%%%%%%%%%%%%%%%%%%%%%%%%%%%%%%%%%%%

\clearpage

%%%%%%%%%%%%%%%%%%%%%%%%%%%%%%%%%%%%%%%%%%%%%%%%%%%%%%%%%%%%%%%%%%%%%%%%%%%%%%%%

\clearpage

%%%%%%%%%%%%%%%%%%%%%%%%%%%%%%%%%%%%%%%%%%%%%%%%%%%%%%%%%%%%%%%%%%%%%%%%%%%%%%%%

\clearpage

%%%%%%%%%%%%%%%%%%%%%%%%%%%%%%%%%%%%%%%%%%%%%%%%%%%%%%%%%%%%%%%%%%%%%%%%%%%%%%%%

%%%%%%%%%%%%%%%%%%%%%%%%%%%%%%%%%%%%%%%%%%%%%%%%%%%%%%%%%%%%%%%%%%%%%%%%%%%%%%%%

\end{document}